# INTERSTITIAL CELLS AND NEURONS RESPOND TO VARIATIONS IN HYDRATION

*Maria P. McGee[1][*], Michael Morykwas[1], Mary Kearns[1], Anirudh Vashisht [1,3], Ashok N. Hegde[2,3] and Louis Argenta[1]*

[1]Department of Plastic and Reconstructive Surgery & [2]Department of Neurobiology and Anatomy. Wake Forest University Medical School, Medical Center Boulevard, Winston-Salem, NC 27157.

[3]Department of Biological and Environmental Sciences, Georgia College and State University, Milledgeville, GA 31061.

*Corresponding author

Maria P. McGee

Dept. Plastic and Reconstructive Surgery

Wake Forest University Medical School

Medical Center Boulevard, Winston-Salem, NC 27157

Phone (336) 716-6716

Fax (336) E-mail address: mmcgee@wakehealth.edu

**ABTRACT**

Dehydration and brain interstitial fluid alterations associate to cognitive dysfunction. We now explore whether changes in matrix hydration are a possible common signal for modulation of water transfer-rates and neuron function.

Hydration levels were determined in brain explants by measuring fluid-transfer rates under controlled gradients of water chemical potential. Transfer-rates changed with the imposed gradient giving *hydration potential* (HP) equivalent to sub-atmospheric pressures ranging 29-103 mmHg in outbred pigs and 59-83 mmHg in inbred mice. This potential increased by approximately 40% after mild dehydration *in vivo*. Brain cells *in vitro* responded to hydration variations within these ranges. Interstitial cells including astrocytes, pericytes and fibroblasts accelerated fluid efflux from three-dimensional collagen matrices at the higher hydration levels. In contrast, electrically stimulated hippocampal neurons increased CREB (cAMP-responsive element binding) phosphorylation at the lower hydration levels.

Results point to a possible link between modulation of interstitial hydration by contractile cells and neuronal responses. Brain tissue *ex vivo* generated suction forces that suggest commensurate counterbalancing forces *in vivo*. These may be provided in part by interstitial cells as they align their mechanical action to regulate local hydration, either accelerating or slowing water flux across overhydrated or dehydrated matrices. Modulation of interstitial hydration would also impact the effective concentration and transfer rate of ions, shed ectodomains and neurotransmitters. Better understanding of brain interstitial fluid control may help advance research in fluid exchange, brain oscillatory activity, drug delivery, clearance of metabolites, and the cognitive dysfunctions of dehydration and aging.

# 1.INTRODUCTION

Dehydration of the brain interstitial matrix leads to neuronal dysfunctions by mechanisms that are not fully understood [1-5]. Matrix hydration/dehydration shifts imply changes in the water activity and necessarily, in the activity of all extracellular solutes; generally, the effective density of all hydrophilic molecules and surfaces increasing with dehydration [6]. Interstitial cells appear to monitor and regulate the local microenvironment of neurons presumably to optimize their responses [7-11], but it is not known whether brain interstitial cells can sense water activity changes directly. Here we measure hydration variations in brain tissue, and responses of interstitial cells and neurons to these variations *in vitro*.

The local water activity of biological matrices *in vivo* is a function not only of their water content but also their specific composition and fluid dynamics and these vary among tissues and with time within each tissue. The total volume of water moving to and from the brain tissue is very large, in rodents estimated at ~ 720 ml /day/g and, at more than ~1200 ml /day/g in humans [12, 13]. Despite the massive volumes exchanged and the importance of water activity in determining local nutrients, metabolites and transmitters concentration, many aspects of water transfer regulation and its pathologies in the brain matrix remain speculative [12-18].

The structural arrangements and composition of interstitial spaces in brain tissue present both similarities and differences with other connective tissues. As in other organs, it includes fibroblasts embedded in loose collagenous tissue surrounding penetrating vessels and pericytes associated to the basal lamina of micro vessels and capillaries [17-21]. In contrast, the inter-neuronal spaces in the neuropil contain little collagen but are rich in glycosaminoglycans including hyaluronic acid and various proteoglycans structures [22-26] known to have very large hydrodynamic volumes and non-ideal osmotic behavior *in vitro*; the osmotic pressure of their solutions increasing disproportionally with concentration [22]. The main cell types occupying the interstitial spaces between the exchange vasculature and neurons are astrocytes and microglia rather than fibroblasts, and appear strategically organized to actively optimize tissue-blood exchange and local transfer of water and solutes at the spatial and temporal scales of neuronal activity [17-19]. Recent evidence from super-resolution brain imaging is consistent with the existence of water fluxes at such scales [27].

A large proportion (78-90 %) [13] of the water in the capillary is exchanged between blood and brain tissue in a single passage, presumably by simple diffusion [28-32]. Hydrostatic and oncotic pressure gradients between the capillaries and the interstitium as well as the fluid resistance and tortuosity of molecular paths can in principle modulate water flux across tissue matrices and possibly across paravascular drainage spaces [28-30], beyond steady state diffusional exchange [31-33]. In general, directly measured interstitial fluid hydrostatic pressures in various tissues under physiologic conditions are reported to fluctuate above and below atmospheric pressure within a range of only a few mmHg [34-36]. The relationship among the relevant pressure differences and local structural components is classically described by equations of the type first attributed to Starling [36] where the local matrix contribution is lumped with the resultant of hydrostatic and colloidosmotic pressures in the interstitial fluid. This resultant does not address the intrinsic pressure and fluid conductance of the extracellular matrix gel nor the possible roles of interstitial cells in local water regulation. We developed the concept of hydration potential (HP) to empirically evaluate matrix forces in the absence of capillary pressures while maintaining the tissue structure and composition [37-40]. The existence of a net sub-atmospheric pressure in most tissues has long been known and is demonstrated by the observation that explants isolated from the blood circulation swell in physiologic solutions at ambient pressure. Although there is little quantitative data for brain, in other tissues this, suction, imbibition or swelling pressure seem to be much higher than the colloidosmotic pressure measured in interstitial fluid [22, 34-36, 41].

In the present study we explore brain matrix hydration potential and cell responses to water activity gradients within the ranges experienced *in vivo* during dehydration and aging. We measure fluid transfer power in isolated brain fibroblasts, pericytes and astrocytes and phosphorylated CREB (cAMP response element) expression [42] in electrically stimulated mouse pyramidal neurons. Together, the results supply evidence for hydration gradients in the brain matrix and for the ability of interstitial and parenchymal cells to sense and respond to these gradients.

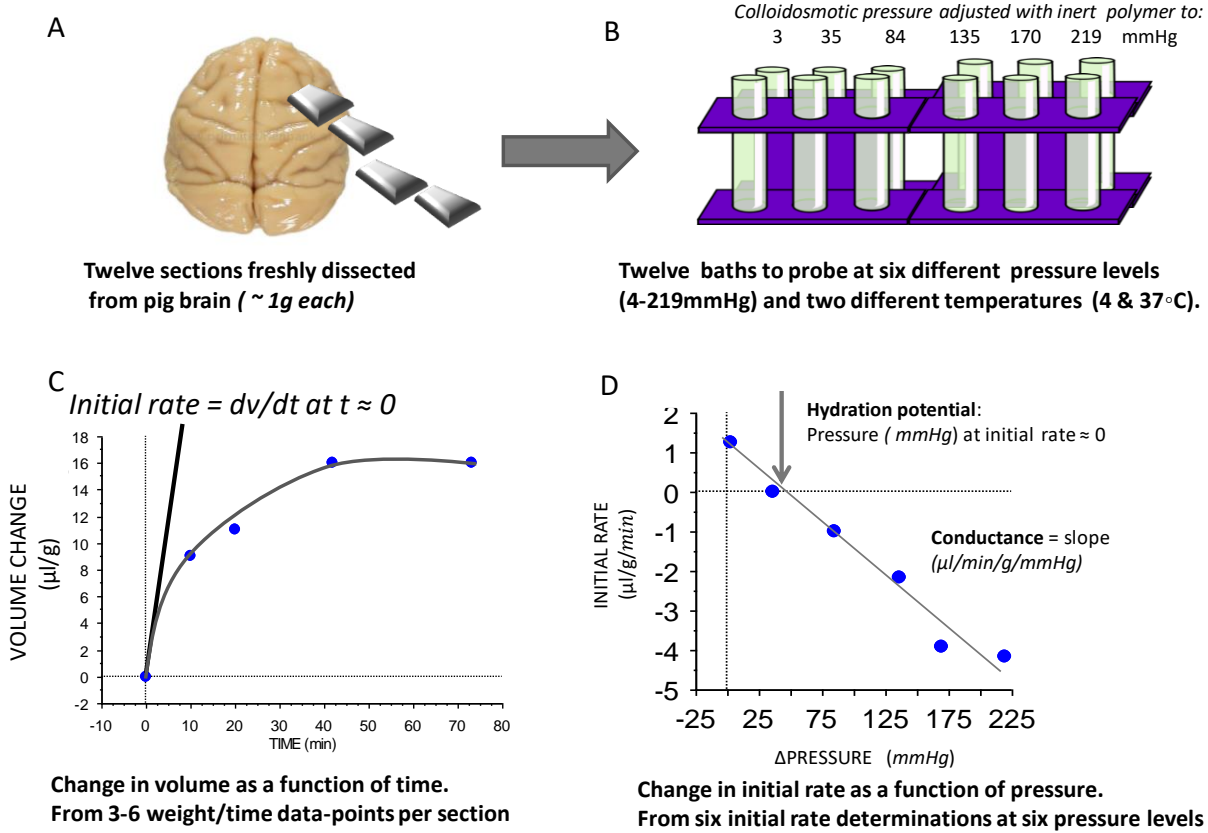

FIGURE 1

**Protocol for determining hydration parameters of brain matrix**.

A. Brains are sectioned to obtain explants that include approximately an equal volume of white and grey matter. **B**. The explants are randomized and immersed in baths of physiologic solution at either 37°C or 4 °C and at 6 different water activity levels for each temperature. **C**. The rate of water efflux/influx from/to the explants in each bath is determined from the change in explant weight with time. **D.** The initial rates of influx/efflux are then plotted against the bath colloidosmotic pressure and straight lines fitted to data-points to obtain the **conductance** from the slope and the **hydration potential** from the rate/pressure intercept at the point where the rate = 0.

## 2.RESULTS

### 2.1 Brain tissue explants respond to $\Delta\mu_w$ variations.

Brain tissue samples when immersed in a set of replicated baths differing only in $\mu_w$, either imbibed or expelled water depending on $\mu_w$ of the bathing fluid. (FIG.2A) The $\mu_w$ in the physiologic baths was reduced by increasing the concentration of the inert polymer polyethylene glycol (~ 5 nm diameter) adjusted to give colloidosmotic pressure increments ranging from 4 to 200 mmHg. This range includes colloidosmotic pressures below and above plasma and a corresponding decrease in $\mu_w$ relative to physiologic and artificial cerebrospinal fluid devoid of proteins. The change in the rate of water-transfer to or from the tissue was directly proportional to the pressure changes and proportionality constants derived from the slope of plots of initial transfer rates versus pressure with units in, $\mu l/min/mmHg/g$ reflect the characteristic hydraulic conductance and fluid transfer power of the brain interstitial matrix (FIG.2B). The hydration potential of the excised brain tissue was calculated from the bath colloidosmotic pressure at which no fluid was transferred to or from the tissue when the water potential in the brain tissue match that in the bath solution; it has the units of a pressure difference (relative to control isotonic physiologic solution) and has values 64 ± 8.1 and 79.6 ± 10.4 *mmHg* ant 37 and 4 °C, respectively. The conductance, was 0.029 and 0.013 $\mu l/min/mmHg/g$ at 37 and 4 °C. The brain tissue conductance (*p-value < .0001*) but no its hydration potential *(p-value = 0.17)* varied significantly with the bath temperature; there was a ~2 fold decrease in the conductance when the temperature was lowered from 37 to 4 °C (FIG. 2C).

Compared to parameters in the other tissues, the brain hydration potential was somewhat higher than in myocardia at 37 °C (*p-value = 0.0013*) and lower than in dermis at 4 °C *(p-value = 0.0163)*. The temperature dependence of brain tissue hydration parameters differed from that of myocardial and dermal explants; in both these tissues, the conductance did not change but the hydration potential increased significantly when the temperature was lowered from 37 to 4 °C *(p-values < 0.001)*. Results from these comparisons are summarized in FIGURES 2 B and C and further discussed in Appendix 2.

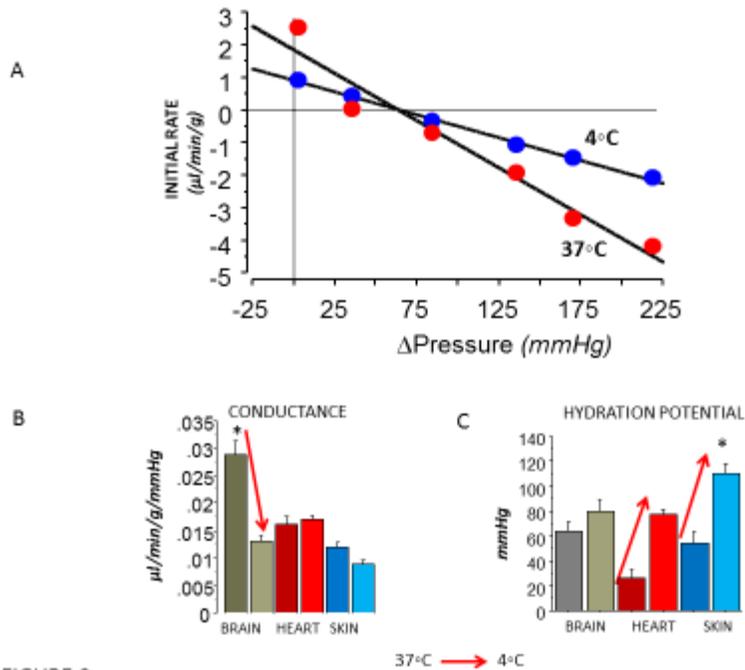

Figure 2. **Interstitial fluid transfer parameters in brain compared to heart and skin.**

**A**. Initial rate of water influx/efflux in brain explants measured at either 4 or 37 °C are plotted as a function of the bath colloidosmotic pressure and fitted with straight lines ($r^2 \leq 0.95$). Each data point is the mean of at least 10 rate determinations. **B**, *Conductance* and **C**, *Hydration potential* values in brain, heart and skin. The arrows indicate statistically significant differences (ANOVA and Fisher's PLSD, *p-values < 0.05*, n = 11, 10, and 8 for brain, heart and skin, respectively) with temperature within each organ, while * indicates differences among organs.

*2.2 Changes in brain hydration parameters after mild dehydration in vivo.*

The experiments in outbred swine included animals under protocols not controlled for water intake. The experimental variability observed may also reflect the genetics of outbred pigs and differences in systemic hydration secondary to each animal diverse history. To further examine physiologic and experimental variations in brain matrix hydration parameters, inbred mice 6-8 weeks of age were either given free access to water or deprived of water for either 12 or 24 hours, but otherwise maintained under identical conditions. In the inbred mice with free access to water, the parameters determined at 25 °C were 73.8 ± 2.8 mmHg for hydration potential and 0.082 ± 0.006 µl/min/mmHg/g for conductance. Water deprivation increased the brain hydration potential by 18 and 24 mmHg (14 and 40%) after 12 and 24 hour dehydration respectively, without significantly changing its fluid conductance. The mean parameters values were similar to those in outbred pigs, but their variations were significantly smaller. The mild water deprivation protocol implemented here did not cause visible distress to the mice which remained alert and active, behaving indistinguishably from the animals that were not deprived. Previous studies in injured tissues from outbred pigs including measurements in experimental dermal burns and edematous myocardia after ischemia reperfusion [37, 51], indicate that the local hydration potential values after injury *in vivo* can vary > 100 mmHg above and below the physiologic levels.

*2.3 Collagen gel contraction by fibroblasts, astrocytes and pericytes as a function of water chemical potential.*

The observed temperature effects on the hydration parameters (Section 3.1 and Appendix 2) suggest that the water-transfer energy measured in the tissue explants may be modulated by metabolically active cells. This type of modulatory action would require the cells to detect small water activity gradients in their microenvironment and respond mechanically to readjust hydration. To test this possibility, mechanical responses of three primary cell lines isolated from human brain interstitium were measured at four different water potential levels.

The three phenotypically distinct interstitial cell types embedded into 3D collagen gels reduced the size of the gels in a time dependent manner. The size of gels either without cells or with heat-killed cells did not change (FIG 3A). Changes in the size of the cell-loaded gels were detectable after variable time lags. For each cell type, the apparent time lags, gel-contraction rates and volume of fluid expelled from the gels changed in proportion to the change in water potential. The tested range of water potential variation corresponds to increments of 4, 42, 100 and 200 mmHg in the colloidosmotic pressure of the bulk media hydrating the gels (FIG 3B).These changes are very small fractions ~ 0.00068, 0.007, 0.017 and 0.03, respectively, of the average physiologic osmotic pressure in human plasma (Appendix1.3 ).

These data show that the cells detected differences as little as ~ 40 mmHg corresponding to ~ 1/150 of plasma osmotic pressure or approximately a 2 milliosmole change. The sensitivity was highest in gels with pericytes giving significantly steeper slopes in rate versus pressure plots than astrocytes and fibroblasts. Contraction rates were significantly slower with astrocytes than fibroblasts and pericytes, while time-lags were shortest with fibroblasts (FIG. 3 C, D, E)

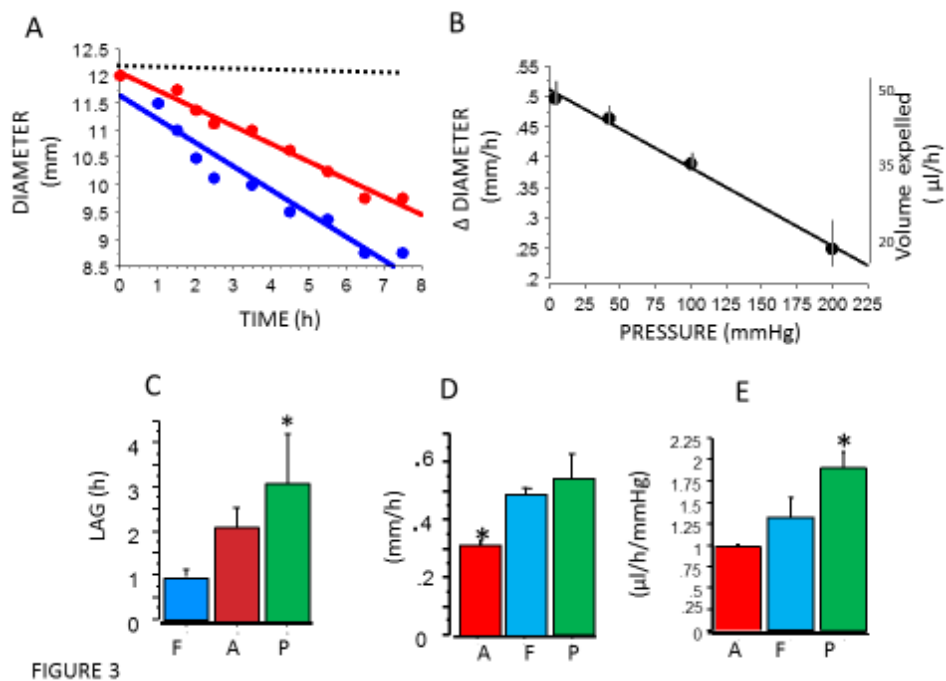

Figure 3. **Fluid transfer from three-dimensional collagen gels mediated by brain interstitial cells responding to water activity.**

Fibroblasts, astrocytes and pericytes were incorporated (7.5-13 x $10^4$ / 500 µl) into collagen gels at water-activity levels expanding the hydration potential measured in brain explants. Cells were incubated at 37 °C in 13 *mm* diameter plastic wells and the water activity in the 3D gel cultures was reduced by adding inert polymer as co-solute to increment the colloidosmotic pressure in the media from 4 to 200 mmHg. After approximately 24 hours, gels were detached from the wells and changes in dimension of the floating gels determined at 30-60 min intervals. Panels A and B illustrate typical time and pressure relationships using results with fibroblasts and in panels C (lag) D (rate) and E (power), the responses of fibroblasts (F, blue) astrocytes (A, red) and pericytes (P, green) are compared. Bar grafts represents means and standard errors, of *n = 4* experimental sets each in duplicate. Statistical significance based in ANOVA and PLSD with *p-values ≤ 0.005*.

A. Gel diameter changes over time only when seeded with live cells. Data points and fitted lines illustrate time-trajectories obtained from a typical collagen-gel contraction assay with brain interstitial fibroblasts. The dimension of gels without cells or with heat-killed cells did not change with either time or the water-activity (dotted line), while the diameter of gels seeded with fibroblasts decreased with time. Figure represents trajectories from gels at 4 and 100 mmHg, blue and red line, respectively.

B. Brain interstitial cells detect and respond to water activity changes. Rates of diameter change calculated from the slope of straight lines fitted to diameter/ pressure data points. The right scale gives the corresponding changes in the volume of fluid expelled from the gels. Reducing the water activity of the media with inert solute proportionally reduced the contractile power exerted by the cells on the gel.

C. Interstitial cell- types differ in their response time. Time elapsed until a change in gel dimensions is first detected is plotted as a function of colloidosmotic pressure in the solution. Contractile responses are delayed but the observed lags were significantly longer with pericytes than with fibroblasts and astrocytes.

D. <u>Interstitial cell-types differ in their contraction rate</u>. The linear rate of change in diameter of cell-seeded gels was significantly lower with astrocytes than with fibroblasts and pericytes. Rates shown are at 4 mmHg.

E. <u>Interstitial cell-types differ in their sensitivity to water activity</u>. Water efflux from gels changed linearly with the colloidosmotic pressure and volume-efflux per unit of pressure was significantly higher in pericytes than astrocytes and fibroblasts

.

The shape of the three cells types in the 3D gels appeared fusiform in general. However, the actual shape was no discernible in any one plane because the cell body and cytoplasmic projections extended irregularly over several slices in the stack. Although this precluded reliable measurements of any potential changes in cell size from the 2D images, nuclear shapes appeared oval, and were measurable in all three cell types. The largest nuclear areas were in cultures with higher water activity (Δ 4 mmHg) where cells populations applied stronger contractile forces (FIG. 4)

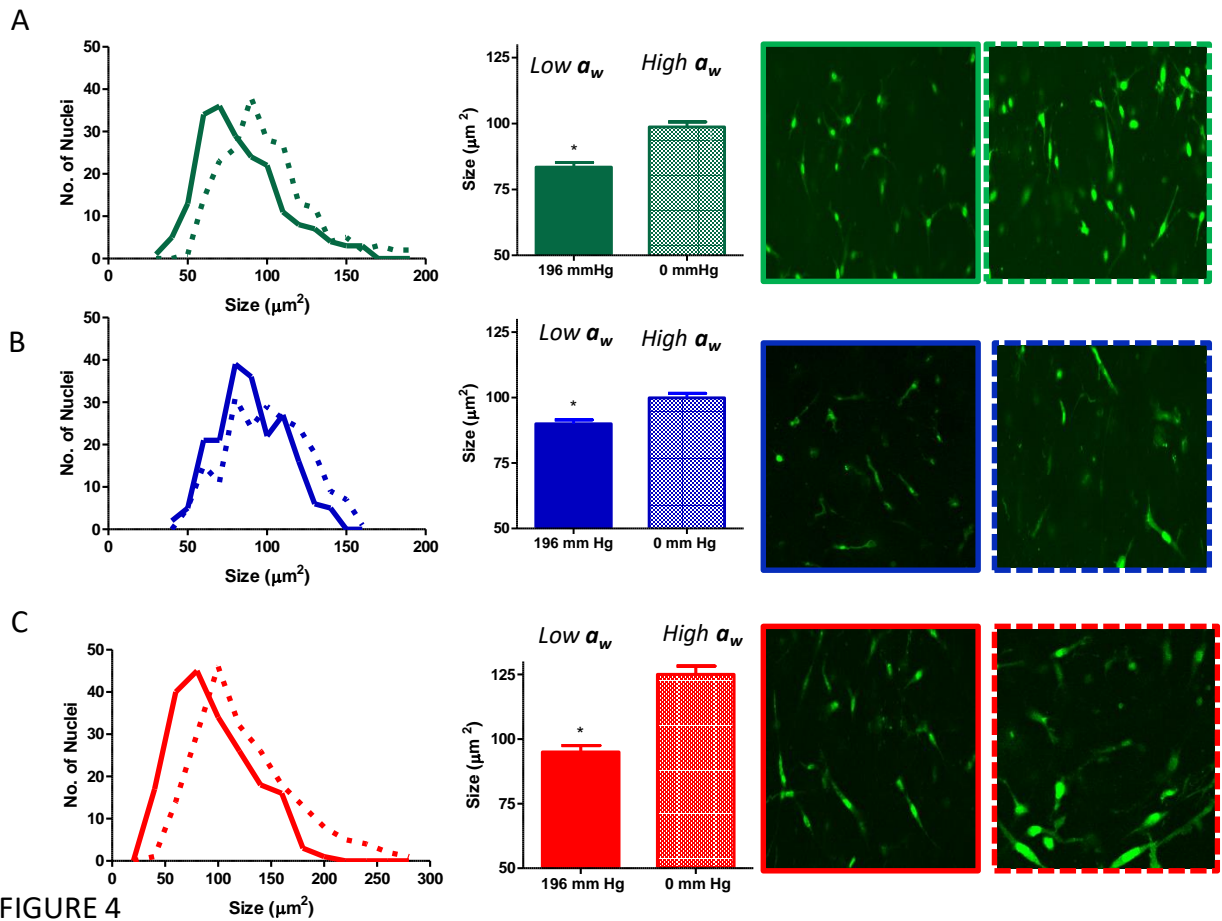

FIGURE 4

Figure 4. **The nuclear size of brain interstitial cells correlates with contraction rate.**

Cells incorporated in 3D collagen gels were immunolabelled with anti f-actin antibodies and Sytox green and examined using confocal microscopy. Nucleus area (measured from the larger nuclear projection) was determined in 2D images of at least

200 cells per gel using Image J software. The distribution of nuclear sizes at high (4 mm Hg dotted line) and low (200 mmHg solid line) water activity is shown for pericytes, fibroblasts and astrocytes in panels a, b, and c, respectively. The corresponding bar graphs d, e, f on the right show the mean and standard errors of nuclear sizes at the high and low water activity levels, contracting fast and slow, respectively. Statistically significant differences (*p-value < 0.05 in t-test*) between the nuclear sizes are indicated by *.

*2.4 Dehydration increases neuronal signaling to the nucleus*. To determine if brain matrix dehydration also influences neuronal responses, we measured plasticity-related signaling in hippocampal slices. Possible changes in the levels of pCREB with water activity were explored in a previously characterized *ex vivo* model of brain plasticity regulation [48, 49].

In slices equilibrated at 196 mmHg the levels of pCREB were increased by nearly 3 fold relative to either unstimulated and stimulated slices equilibrated in artificial cerebrospinal fluid ( FIG.5).The increase was stimulus-dependent and similar in magnitude to that seen in the non-dehydrated positive controls treated with β-lactone and did not increased further when dehydration was combined with β-lactone treatment suggesting that CREB phosphorylation is maximally stimulated under either condition.

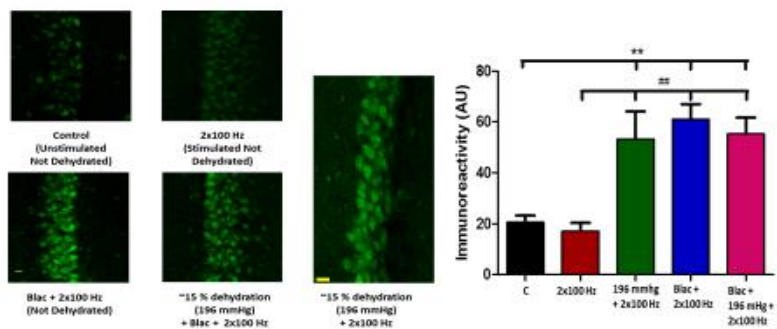

FIGURE 5

FIGURE 5. **Dehydration increases CREB phosphorylation (pCREB)**

<u>Left panels</u>. Confocal images of pCREB immunoreactivity in the CA1 region of hippocampal slices not stimulated (negative control); stimulated by a sub-threshold LTP induction protocol (2X 100 Hz) and treated with β-lactone (Blac positive control); stimulated not dehydrated ; Dehydrated (196 mmHg) and β-lactone treatment; dehydrated and stimulated. Scale bars 20µm. <u>Right panels</u>. Quantification of pCREB immunoreactivities shows that dehydration significantly increases CREB phosphorylation to levels observed with β-lactone treatment and that the β–lactone and dehydration effects are not additive. Comparisons (one way ANOVA) as indicated by horizontal lines ** and ##; $p < 0.001$.

# 3. DISCUSSION

## 3.1. *Variations in brain HP ex-vivo suggest water activity gradients in vivo*

The present studies show that the average HP in freshly isolated brain tissue explants from swine and mice is higher than in cerebrospinal fluid and plasma. During equilibration *ex-vivo* the tissue explants varied their hydration levels by suctioning or expelling water. While equilibrating against osmotic pressures exceeding that of cerebrospinal fluid and plasma - by up to ~ 60 and 30 mmHg, respectively- water transferred passively into the brain tissue. Water influx against these pressures indicates that the brain matrix is dehydrated relative to cerebrospinal fluid and plasma. The empirically obtained value for the HP in the pig brain is of the same order than that in myocardial, dermal and mouse brain tissue, in the range of systolic- diastolic pressure differences and twice as large as plasma colloidosmotic pressure. This suggest colloidosmotic gradients that are of reverse direction to those customarily predicted from the low protein content of cerebrospinal fluid.

The similar value of HP variations observed among different tissue matrices and animal species also suggests that these values represent physiologic ranges for matrix hydration. Significant changes in the matrix suction potential were demonstrated after mild dehydration in mice, in the dermis and myocardia during interstitial edema, [37, 51] and in brain during age-related dehydration [4] indicating that even larger variations are possible during pathological conditions.

## 3.2 *Significance of interstitial cells responsiveness to water activity gradients*

*In vivo*, during capillary transit times of ~1-2 second, most of the blood water exchanges by rapid diffusion with the extravascular water in a single passage [12, 13, 32]. The exchange rates may not be constant but perhaps fluctuate with the water activity thus influencing diffusion and volume transmission in the brain [15, 19, 28]. Interstitial and parenchymal cells may also respond to these transient shifts if brain cells can detect differences in water activity. We found that cells isolated from brain interstitium responded to small variations in matrix hydration. As the water activity increased above the physiologic levels found in the brain explants, so did the fluid transfer-work and power of the interstitial cells, *in vitro*. If responses were qualitatively similar in the complex topology of the brain matrix but at the velocities reported for cell motility *in vivo* [52], they would also help counterbalance suction forces by dissipating local gradients at µm/s spatiotemporal scales [19, 23, 27, 53].

## 3.3 *Significance of neuron responsiveness to water activity gradients*

In hippocampal slices (after 1hour recovery), the levels of pCREB increased in electrically stimulated neurons when the water activity decreased. We take these results together with those above (sections *2.1 & 2.2*) as evidence consistent with the idea that water activity is a common variable linking the matrix' material and colligative properties (sections *2.1&2.2*), interstitial cell action (section *2.3*) and neuronal function. In the three different experimental set-ups - tissue explants, interstitial cells in 3D collagen matrices and hippocampal organotypic cultures- responses were detected within ranges of physiologic and pathologic HP variations.

Induction of synaptic plasticity to model memory and cognition *in vitro*, requires activation of the cyclic adenosine monophosphate (cAMP) signaling pathway, leading to gene transcription -via CREB transcription factor- and expression regulated by the proteasome [48]. This is the main transcription factor linking synaptic stimulation with gene expression and is activated by phosphorylation. Phosphorylated CREB (pCREB) induces specific immediate-response genes via cAMP-responsive element (CRE) in their promoter regions. Therefore the increase in pCREB seen here is also in line with previous results indicating an increase in basal synaptic transmission and plasticity dysregulation in brain slices equilibrated at water activity levels below physiologic [4], and with a variety of studies in animal and humans suggesting association between cognitive disturbances and dehydration [1-3, 5]. It is important to note however, that besides synaptic plasticity, CREB influences neuronal responses to other extracellular stimuli and that the processes by which intracellular pathways are modulated by pCREB levels remain largely unknown [42, 50].

*3.4 Possible mechanisms mediating cell responses to water activity gradients*

Under the conditions of these experiments, both interstitial cells and neurons seem capable of evaluating environmental hydration, detect minute changes in water activity, modulate nuclear signaling and in the case of contractile cells, adjust and align their power with that of passive water transfer. This suggest that contractile interstitial cells are programmed to efficiently modulate hydration; working within other imposed constraints- such as physiologic pH, temperature, tensional forces, brain topology and the genetically encoded metabolic pathways –they would tend to minimize the overall energy expended by summing their mechanical action to passive transfer along water activity gradients.

Within the range of water activity gradients tested, there were no apparent mechanical signals such as gel swelling or changes in gel stiffness, tension or shear since the cultures were static and hydration was controlled with inert soluble polymers that fully penetrate the collagen mesh and the fluid was expelled exclusively by the coordinated mechanical action of the cell population embedded in the gel. These observations raise the question of how cells detect and respond to minor changes in the water activity at these mesoscopic scales. Changes in water activity necessarily determine changes in effective concentration of all solutes and there is a link between the hydration state of molecules and their conformation. The rate of functional protein conformational transitions in solution is linked to water transfer [44, 53]. As macromolecular ensembles shift the range of their dynamic fluctuations to populate the more stable conformations compatible with each water activity level, lower water activity favors compact, relatively dehydrated conformations, while higher water activity favors open, more hydrated conformations [44, 54-59].

At interfaces between the cell and the interstitial matrix the molecular adjustments to water activity must therefore be global including all molecules and structures, but we could predict that molecules with the structural motives most sensitive to hydration/dehydration conformation shifts could modulate signaling. With this in view, it is therefore interesting that many of the better studied signaling processes between matrix structures and cells as well as during synaptic plasticity require inter and intra molecular water transfer. The functional conformational transitions of integrins and AMPA and NMDA receptors involve changes in inter and intra molecular solvent accessible spaces [54, 55]. For example, the open active integrin conformation has much larger solvent volumes in the inter subunit space compared to the compact low affinity conformation. For the glutamate receptors, interfaces between dimers in the desensitized structures appear to separate, also increasing the intermolecular solvent spaces relative to activated conformations. Although structure-function correlations of macromolecules are very complex it is clear from available information that functional transitions associate to structural sifts linked to transfer of large volumes of water to or from the protein surfaces and/or water accessible pockets. Water transfer cost energy that is reduced (or increased) in the presence of water activity gradients. For example, the rates of -glycosaminoglycan-protein interactions during activation of antithrombin and coagulation proteases are accelerated (and the free energy of activation reduced) as the water activity decreases [56, 57]. It is to be expected that modulatory effects of water activity gradients are pronounced in crowded biological microenvironments [58] such as the brain intra and extracellular matrix spaces where competition for hydration water is the rule.

It is also possible that the hydration state of certain supramolecular structures such as peri-neuronal nets, membrane proteoglycans, and paravascular collagen interfaces translate into localized changes of the matrix' material properties that the cells "sense". Cells nevertheless seem to detect and respond to water activity changes in the absence of any apparent mechanical perturbations transmitted by a gelled structure as exemplified in APPENDIX 1. The proliferation of cells suspended in liquid aqueous media devoid of collagen or glycosaminoglycans, was exquisitely sensitive to, and linearly dependent on, the water chemical potential but essentially independent of the chemical configuration and solution properties of the inert solute used to control it.

### 3.5 *Summary conclusions and perspectives.*

These results in brain interstitial matrix and cells are consistent with the concept that at every scale, the brain continually senses and respond to minute hydration changes. According to this view, at the molecular level, conformations that mutually balance acceleration and retardation steps in signaling pathways are appropriately shifted by hydration/dehydration transitions to regulate the magnitude and/ or duration of cell responses. The contractile cell action appears to have evolved in alignment with passive diffusion and convective transfer mechanisms to modulate interstitial matrix hydration. Whether the hyperexcitability [4, 61] and other neuronal changes induced by dehydration are directly related to memory and learning programs, serve physiologic compensatory functions or reflect a pathology are issues that deserve further investigation [2-5, 14-19, 28]. Fluctuations in water activity may also influence the distribution and local concentration of neurotransmitters shed ectodomains and their spatiotemporal interactions with the matrix [19, 50, 59, 60, 61]. Recent advances in second-harmonic imaging can already detect ion channel activity linked to fluctuating membrane hydration in response to changes in bulk solution's water activity and ionic strength [63]. In a near future, these and other imaging developments [27] will evaluate the impact of water activity fluctuations and their induced water fluxes on cell function at sub-µm and sub-second scales.

From a more direct translational point of view, the findings presented here can help improve the models on which the implementation of new imaging modalities and drug delivery systems depends. Matrix suction-pressures of the magnitude found here would model a more resilient water exchange mechanism than what is possible with the small pressures usually considered in the literature.

## 4. EXPERIMENTAL PROCEDURES

*4.1 Large animal tissue explants.* The brains of 13 female swine (25 kg) were obtained immediately after euthanasia from animals in unrelated study protocols approved by the Institutional Animal Care and Use Committee. The cerebrals hemispheres were separated and the midbrain structures and part of the white matter discarded. Sagittal wedges 1 cm thick were further sectioned to obtain 12 pieces that by visual inspection were approximately 1 cm$^3$ each and appear to include similar volumes of gray and white matter; the actual samples weighed 1010.8 ± 27.1 mg (*mean and SE, n = 149*). To prevent dehydration and/or hydration of samples during preparation, the tissue was wrapped in parafilm® and the time elapsed from brain extraction to initiation of water activity titration experiments was kept within 15-30 min.

*4.2 In vivo brain dehydration. Water deprivation protocols in mice.* Mice were obtained from Charles River (Wilmington, MA) and used according to mild water deprivation protocols approved by the Institutional Animal Care and Use Committee of Wake Forest University Health Sciences. Mice 6-12-week old mice were water deprived for either 12 (n = 5) or 24 hours (n = 5) and their brains removed immediately after euthanasia. Medula and spinal cord connections were excised, and the remaining cerebrum was divided in four approximately equal sections weighing 70.8 ± 1.3 mg *(mean and SE, n = 80)* and by inspection, each appeared to include similar volumes of grey and white matter. During the water deprivation intervals, animals were maintained in the laboratory under observation. Control mice (n = 10) were maintained under identical conditions except that they were allow free access to water. To measure the hydration potential, the protocol was as described for pig brains with the modifications noted in section 4.3.

*4.3 Water activity titrations.* Classical osmotic stress techniques [43, 44] developed to study intermolecular hydration forces and hydration/dehydration transitions are here adapted to derive hydration parameters in brain tissue explants closely following water potential titration protocols previously validated in dermal and myocardial explants [40, 45]. Briefly, to determine the hydration potential and hydraulic conductance of the pig brain matrix, each of 12 brain sections is immersed in a bath where the water activity is experimentally controlled by including an inert polymer in an otherwise physiologic electrolyte solution with baseline water potential of ~5800 mmHg. The water potential in the baths is adjusted by addition of the inert colloids at increments ranging from 4-219 mmHg in a volume > 50fold in excess of the interstitial fluid volume in the explants. In the experiments using explants from pig brains, the temperature of the bath is maintained constant at either 4 or 37°C and water potential in each of the six baths maintained at one of six different levels. In the experiments with explants from the much smaller mouse brains, four tissue sections were immersed in artificial cerebrospinal fluid containing 125 mM NaCl, 3 mM KCl, 2.3 mM CaCl, 1.3mM MgCl$_2$, 25mM

NaHCO$_3$, 1.25 mM NaH$_2$ PO$_4$ and 10 mM Glucose (pH 7.4) and the temperature was maintained at 25°C [4].

Upon immersion in the bath the brain explants experience a water driving gradient when the water chemical potential in the bath solution and the brain matrix differ. Each incremental amount of added macromolecular solute decreases the water chemical potential, $\Delta\mu_w$, and the activity, $a_w$ and increases the osmotic pressure, $\pi$ of the bath solution. Variations in either one of these three functions can be used to define changes in hydration; they are closely related:

$\Delta\mu_w = RT \ln a_w = -V_w \pi$                                                    [1]

Where $V_w$ is the molar volume of water and $R$ (8.314 J/mol °K) the gas constant and $T$ absolute temperature, respectively [46].

Considered as an open thermodynamic system, the work of fluid transfer to and from the live brain explants in the bath also includes forces arising from its structural and biological complexity. We previously used the term *hydration potential* to designate the net fluid-transfer forces measured in the interstitial matrix of excised tissues and evaluated it as a thermodynamic grand potential that includes not only water chemical potential but also entropic, interfacial and mechanical components of the fluid-transfer energy [40, 47].

The volume of water transferred is measured as a function of time by precision weighing of the tissue every 5 min for mice brain and every 15 min for pig brain up to 60 min incubation time. The transfer velocity either influx or efflux is calculated from the initial changes in tissue volume with time, $\Delta 1\ mg = \Delta 1\mu l$. Water flows from the tissue to the bath and from bath to the tissue when the potential in the tissue is higher or lower than in the bath, respectively. Under conditions where the transfer velocity is null, the potential in the tissue is assumed to be the same as in the bath. The *hydration potential* gives the excess fluid-transfer energy in the excised tissue relative to physiologic fluid devoid of colloids. The experimental protocol and initial data reduction procedures are outlined in Figure 1.

*4.4 Cells and collagen gel contraction assay*. The primary cultures of human brain interstitial cells used in gel contraction assays were obtained from a commercial supplier (ScienCell Research Laboratories). As per information from the supplier, the fibroblasts were positive for fibronectin, the pericytes for α-smooth muscle actin and the astrocytes for GFPA (glial fibrillary acidic protein).The cells were cultured and replicated as recommended by the supplier in poly-l-lysine-coated flasks and used between passages 2 and 4.

Collagen solutions were prepared at 4 °C with 2.5mg/ml collagen (rat-tail collagen I, Gibco, Life Technologies, Grand Island, NY, USA) and cells were added at the lower density that gave measurable contraction in ≤ 10 hours: this was 7.5 x10$^4$ / 0.5 ml collagen solution, for pericytes and fibroblasts and 13 x 10$^4$ / 0.5 ml collagen solution, for astrocytes. The cells-collagen suspensions were allowed to gel at 37 °C for 1 hour in 13 mm-diameter wells and then the water activity of the gels was adjusted by adding 1 ml of polyethylene glycol containing culture medium to increase the colloidosmotic pressure by 42,

100, and 200 mmHg. Control cultures had only the medium with a baseline colloidosmotic pressure of 2-4 mmHg and mimic the culture conditions generally reported for studies of cell mechanics *in vitro*. However, the water activity in these control cultures is higher than the physiologic conditions *in vivo*, as indicated by titrations of brain tissue explants demonstrating hydration potentials above the physiologic saline solution and cerebrospinal fluid. Therefore, the range of pressures chosen for the gel contraction experiments includes pressures variations of $\Delta 4$ and $\Delta 42$ mmHg, and $\Delta 100$ and $\Delta 200$ mmHg which are approximately 60 and 20 mmHg above, and 36 and 140 mmHg below the measured HP in the brain explants, respectively.

Some experiments also included either gels with no cells or gels with heat-killed cells. The colloidosmotic pressure of control and polyethylene glycol solutions were determined from a set of standard curves constructed by membrane osmometry using a 3kDalton cut-off membrane. Like for the brain explants titrations, the range of pressures tested corresponds to less than 5 % of the osmotic pressure of plasma (~5800 mmHg) and is within the random physiological variation measured among human plasma samples. After 15 hours incubation, the gels were separated from the walls of the culture wells and changes in their diameters measured during 5-24 hours at ½-1hour intervals.

*4.5 Toxicity and specificity controls*. The non-adherent human cell line HL-60 was used to test for possible toxicity and/or specific interactions between the polymer and the cells. These experiments show that the inert polymers used to measure hydration parameters are not toxic and that cell responses to $\Delta\Delta\mu_w$ are independent of the chemical structure of the polymer used as solute. Appendix 1.

*4.6 Transcription factor expression in organotypic hippocampal cultures*. Transverse hippocampus slices (400 µm) were prepared from ~8- 9 weeks old mice (n = 6) using a tissue chopper in oxygenated and chilled artificial cerebrospinal fluid and after 1 hour recovery at 32 °C, field excitatory postsynaptic potentials (fEPSP) were recorded in the CA1 region of the hippocampus using a bipolar electrode to stimulate the Schaffer collateral pathway. During both the recovery and stimulation stages, the slices were equilibrated in artificial cerebrospinal fluid at water potential levels approximately 60-100 mmHg below and 140-100 mmHg above the levels determined in brain tissue explants of control and water-deprived animals. The stimulation intensity was adjusted to give ~35% of maximal fEPSP. Subthreshold stimulation relative to long term potentiation (LTP) protocols was given with 2X 100 Hz trains spaced 5 min apart [48]. The hippocampal slices were fixed in 4% paraformaldehyde immediately after stimulation and processed for identification and quantification of phosphorylated cAMP response element binding protein (pCREB), using immunohistochemistry as previously described [48, 49]. In brief, slices were incubated overnight at 4 °C with polyclonal antibodies against pCREB (1:1000, Cell signaling Technology, Danvers, MA), washed, and incubated with secondary Alexa 488-conjugated goat anti-rabbit antibody (1:300, Invitrogen, Grand Island, NY), for an additional 8 hours at 4 °C. Carl Zeiss LSM510 laser scanning confocal microscope was used to image the fluorescence and ImageJ (National Institutes of Health, Bethesda, MD) software was used to analyze fluorescence intensity in the images. At each water activity level, hippocampal slices that received subthreshold LTP stimulation were compared to their time matched controls for quantification of changes in pCREB expression.

Positive controls included slices incubated for 30 min prior stimulation with 25µM β-lactone to increase pCREB phosphorylation [48, 49].

*4.7 Morphologic analyses.* Possible morphologic changes in brain interstitial cells incorporated in the three-dimensional collagen gels at different $\Delta\boldsymbol{\mu_w}$ levels were investigated in representative experiments. For these studies, replicated gels were processed within 30 min culture time and when contraction rates approached a plateau; this occurred approximately at 7 hours in fibroblast cultures and 24 hours in astrocytes and pericytes. The cells were fixed with 2% glutaraldehyde, the cytoskeleton stained with monoclonal antibody against f-actin and the nucleus highlighted with sytox green [45]. The fixed and stained gels were examined by confocal microscopy (Axiovert 100M Carl Zeiss, Jena, Germany), with objective Fluor 20 x in plane, multitrack mode with 10 µm step size. Two-dimensional images of at least 200 cells from different layers and including regions at the center and periphery of each gel were selected using Zeiss LSM image browser and changes on the nuclear area quantified with Image J (NIH, Bethesda, MD, USA).

*4.8 Data processing and analyses.* To determine hydration parameters, initial rates of fluid transfer were calculated from second degree polynomials fitted to the progress curves. Curves were constructed by tracing tissue volume changes over time during equilibration at each $\mu_w$ level. The change in initial flux rates was then modeled as a linear function of the bath colloidosmotic pressure and the conductance and hydration potential determined from the slope, and the pressure at which the initial influx/efflux rate = 0, respectively, (FIGURE 1 C, D). Linear regression analyses and ANOVA and *post-hoc* tests were applied to determine dependencies and to compare results from the various experimental conditions and among tissues, including data from dermis and myocardia from previous studies. The kinetic analyses of results in dermal samples extend calculations based on equilibrium data [38, 39] and kinetic analyses in myocardial explants have been published before [45]. The experimental protocols for pig dermal and myocardial tissue in the previous studies were identical to the one with pig brain described here allowing for the global analyses of results and comparison of parameters presented in this report.

Contraction of 3D-collagen gels by the interstitial cells was modeled as a linearly decaying process and its rate, *R (mm/h)*, calculated from the reduction in gel diameter, *D* with time. Changes in the volume displaced from the gels by the cell action as a function of pressure were calculated from the difference between the initial volume *(~500 µl)* and volumes at time *t* after initiation of gel contraction in replicated gels equilibrated at 4 different water activity levels ; volumes of the floating gels were estimated from their radius *(r ~ 6 mm)* and height *(h ~ 4.5 mm)* scaled by *($D_{(t)}$/D)* and inserted in the formula for the volume of a regular cylinder *($\pi\ r^2 h$).* The progress-curves tracing the change in gel diameter as a function of time in gel contraction assays were analyzed with Table Curve 2D (Systat Software In. Richmond CA, USA).

Linear regression, analyses of variance (ANOVA) and post-hoc multiple Fisher's protected least-significant difference (PLSD) were implemented using Stat View (SAS Institute, Cary NC, USA).

ACKNOWLEDGEMENTS

This work was supported by Research and Development Funds from the Department of Plastic and Reconstructive Surgery, Wake Forest University Medical School.

FIGURE LEGENDS

**Figure 1. Protocol for determining hydration parameters of brain matrix**.
A. Brains are sectioned to obtain explants that include approximately an equal volume of white and grey matter. **B**. The explants are randomized and immersed in baths of physiologic solution at either 37°C or 4 °C and at 6 different water activity levels for each temperature. **C**. The rate of water efflux/influx from/to the explants in each bath is determined from the change in explant weight with time. **D.** The initial rates of influx/efflux are then plotted against the bath colloidosmotic pressure and straight lines fitted to data-points to obtain the **conductance** from the slope and the **hydration potential** from the rate/pressure intercept at the point where the rate = 0.

Figure 2. **Interstitial fluid transfer parameters in brain compared to heart and skin.**

**A**. Initial rate of water influx/efflux in brain explants measured at either 4 or 37 °C are plotted as a function of the bath colloidosmotic pressure and fitted with straight lines ($r^2 \leq 0.95$). Each data point is the mean of at least 10 rate determinations. **B**, *Conductance* and **C**, *Hydration potential* values in brain, heart and skin. The arrows indicate statistically significant differences (ANOVA and Fisher's PLSD, *p-values < 0.05*, n = 11, 10, and 8 for brain, heart, and skin, respectively) with temperature within each organ, while * indicates differences among organs

Figure 3. **Fluid transfer from three-dimensional collagen gels mediated by brain interstitial cells responding to water activity.**

Fibroblasts, astrocytes and pericytes were incorporated (7.5-13 x $10^4$ / 500 µl) into collagen gels at water-activity levels expanding the hydration potential measured in brain explants. Cells were incubated at 37 ◦C in 13 *mm* diameter plastic wells and the water activity in the 3D gel cultures was reduced by adding inert polymer as co-solute to increment the colloidosmotic pressure in the media from 4 to 200 mmHg. After approximately 24 hours, gels were detached from the wells and changes in dimension of the floating gels determined at 30-60 min intervals. Panels A and B illustrate typical time and pressure relationships using results with fibroblasts and in panels C (lag) D (rate) and E (power), the responses of fibroblasts (F, blue) astrocytes (A, red) and pericytes (P, green) are compared. Bar grafts represents means and standard errors, of *n = 4* experimental sets each in duplicate. Statistical significance based in ANOVA and PLSD with *p-values ≤ 0.005.*

A. <u>Gel diameter changes over time only when seeded with live cells</u>. Data points and fitted lines illustrate time-trajectories obtained from a typical collagen-gel contraction assay with brain interstitial fibroblasts. The dimension of gels without cells or with heat-killed cells did not change with either time or the water-activity (dotted line), while the diameter of gels seeded with fibroblasts decreased with time. Figure represents trajectories from gels at 4 and 100 mmHg, blue and red line, respectively.

B. <u>Brain interstitial cells detect and respond to water activity changes</u>. Rates of diameter change calculated from the slope of straight lines fitted to diameter/ pressure data points. The right scale gives the corresponding changes in the volume of fluid expelled from the gels. Reducing the water activity of the media with inert solute proportionally reduced the contractile power exerted by the cells on the gel.

C. <u>Interstitial cell- types differ in their response times</u>. Times elapsed until a change in gel dimensions is first detected are plotted as a function of colloidosmotic pressure in the solution. Contractile responses are delayed but the observed lags were significantly longer with pericytes than with fibroblasts and astrocytes.

D. <u>Interstitial cell-types differ in their contraction rate</u>. The linear rate of change in diameter of cell-seeded gels was significantly lower with astrocytes than with fibroblasts and pericytes. Rates shown are at 4 mmHg.

E. <u>Interstitial cell-types differ in their sensitivity to water activity</u>. Water efflux from gels changed linearly with the colloidosmotic pressure and volume-efflux per unit of pressure was significantly higher in pericytes than astrocytes and fibroblasts

.

Figure 4. **The nuclear size of brain interstitial cells correlates with contraction rate.**

Cells incorporated in 3D collagen gels were immunolabelled with anti f-actin antibodies and Sytox green and examined using confocal microscopy. Nucleus area (measured from the larger nuclear projection) was determined in 2D images of at least 200 cells per gel using Image J software. The distribution of nuclear sizes at high (4 mm Hg dotted line) and low (200 mmHg solid line) water activity is shown for pericytes, fibroblasts and astrocytes in panels a, b, and c, respectively. The corresponding bar graphs d, e, f on the right show the mean and standard errors of nuclear sizes at the high and low water activity levels, contracting fast and slow, respectively. Statistically significant differences (*p-value < 0.05 in t-test*) between the nuclear sizes are indicated by *.

FIGURE 5. **Dehydration increases CREB phosphorylation (pCREB)**

<u>Left panels</u>. Confocal images of pCREB immunoreactivity in the CA1 region of hippocampal slices (A) not stimulated (negative control); (B) stimulated by a sub-threshold LTP induction protocol (2X 100 Hz) and treated with β-lactone (Blac positive control); (C) stimulated not dehydrated ; (D) Dehydrated (196 mmHg) and β-lactone treatment; (E) dehydrated and stimulated. Scale bars 20µm. <u>Right panels</u>. Quantification of pCREB immunoreactivities shows that dehydration significantly increases CREB phosphorylation to levels like those observed with β-lactone treatment and that the β–lactone and dehydration effects are not additive. Comparisons (one-way ANOVA) as indicated by horizontal lines ** and ##; p <0.001.

**FIGURES**

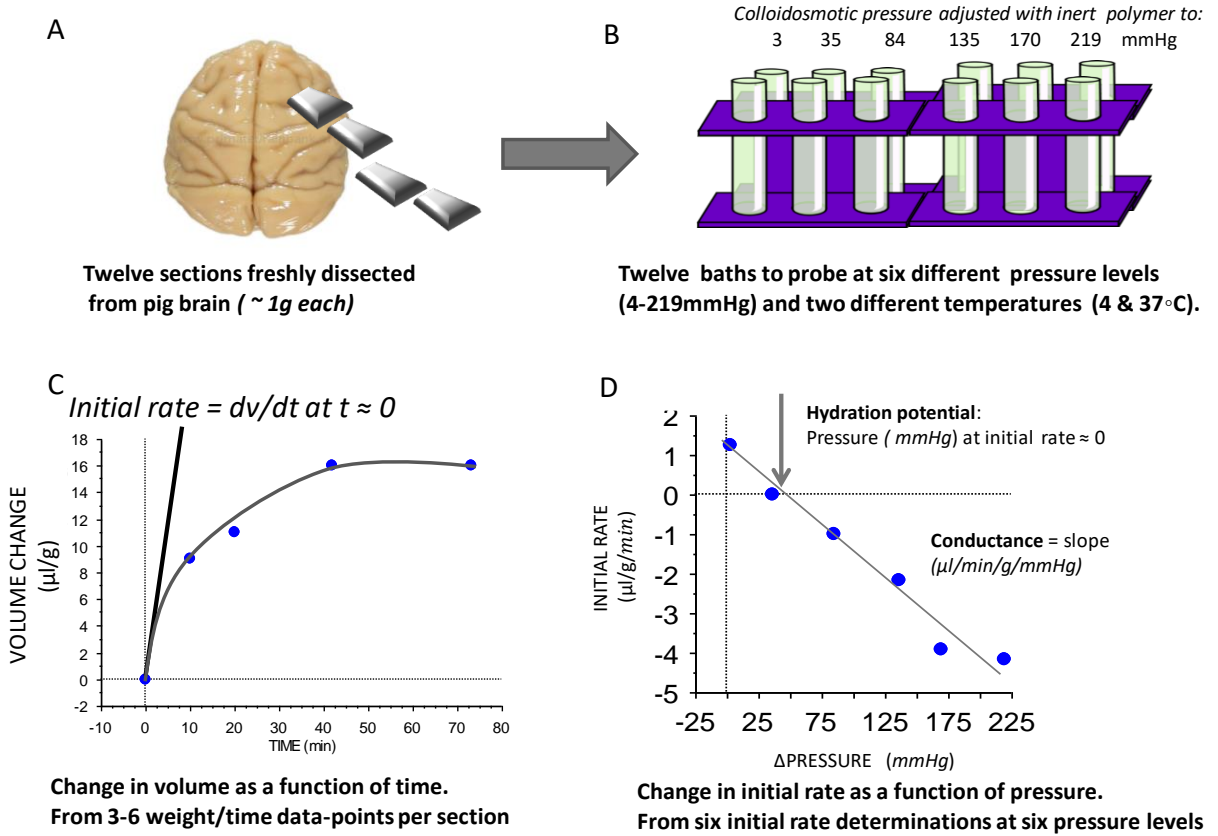

FIGURE 1

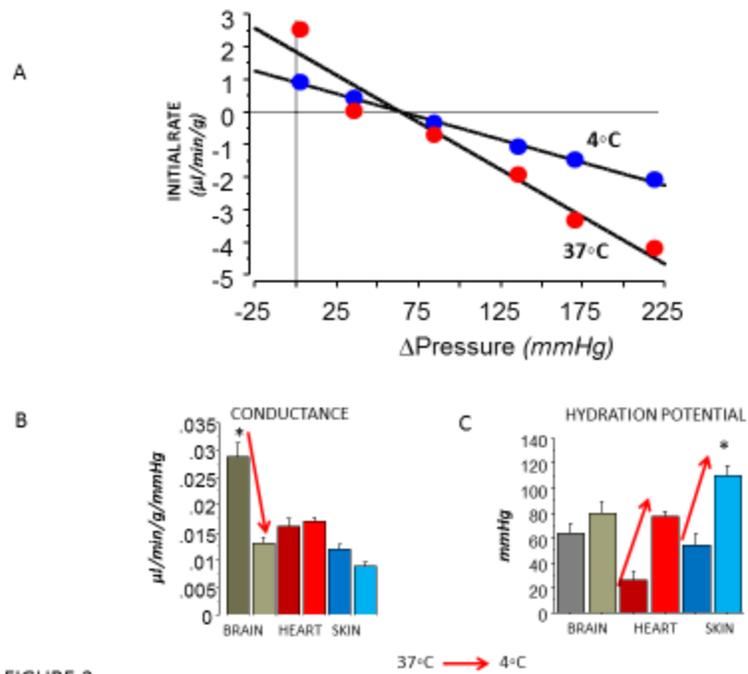

FIGURE 2

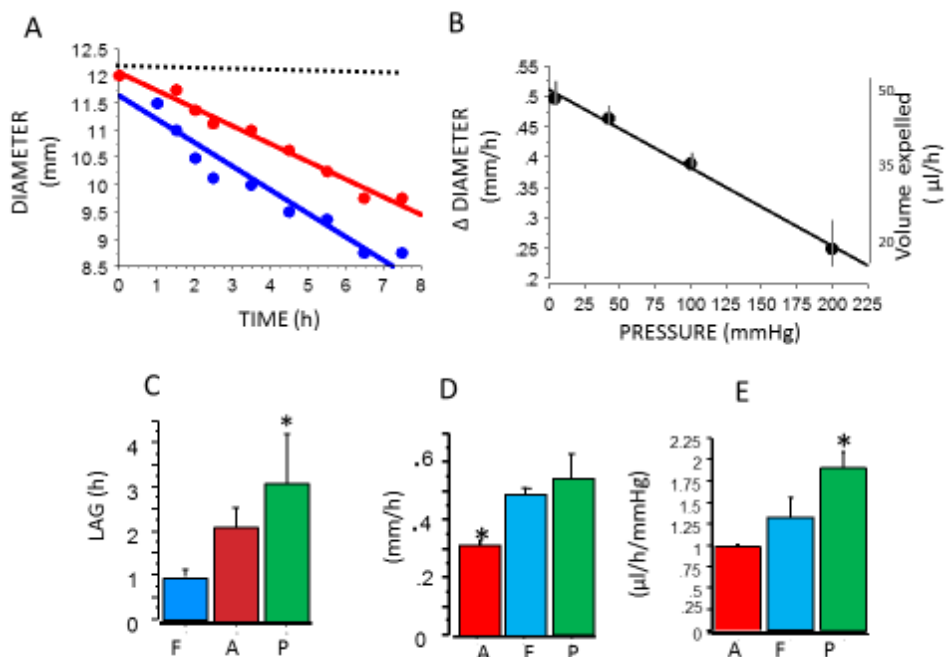

FIGURE 3

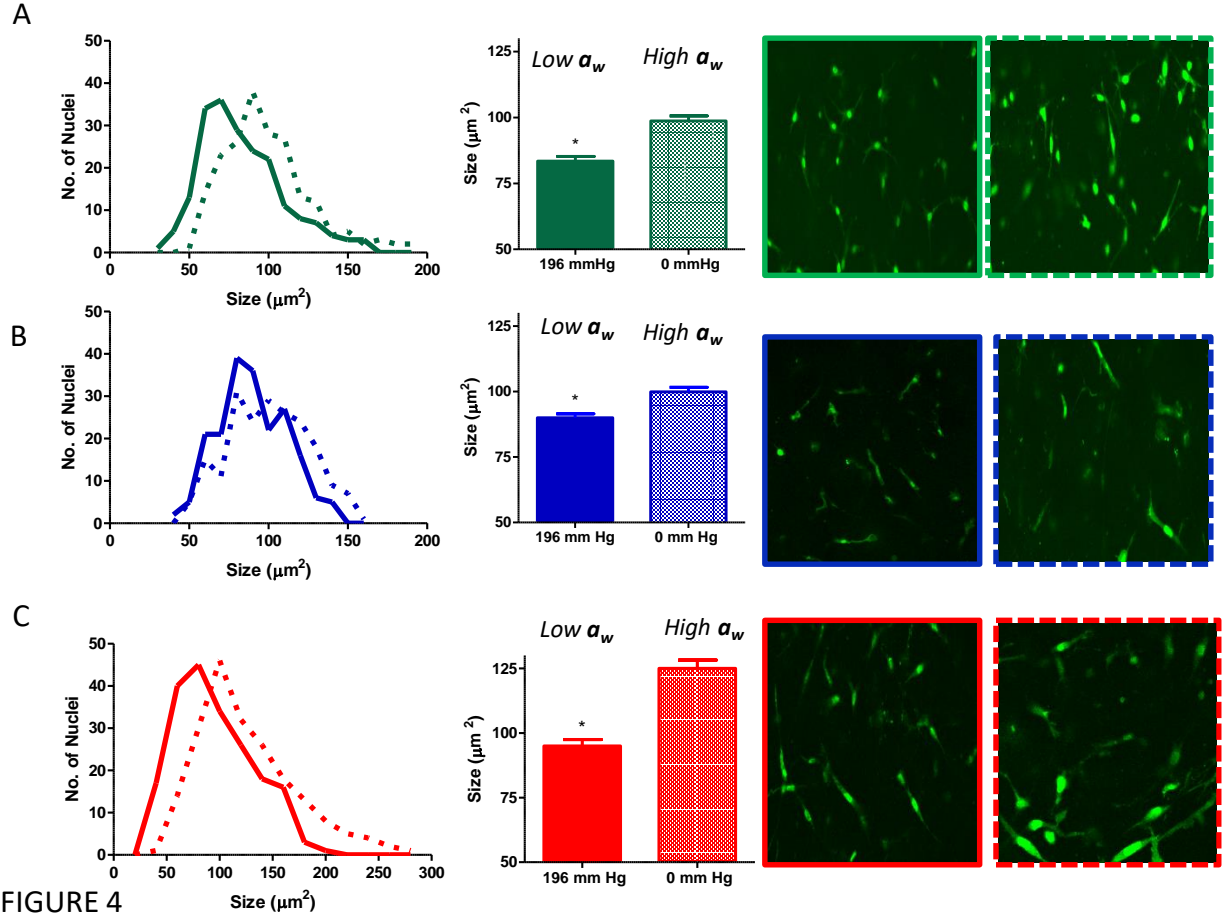

FIGURE 4

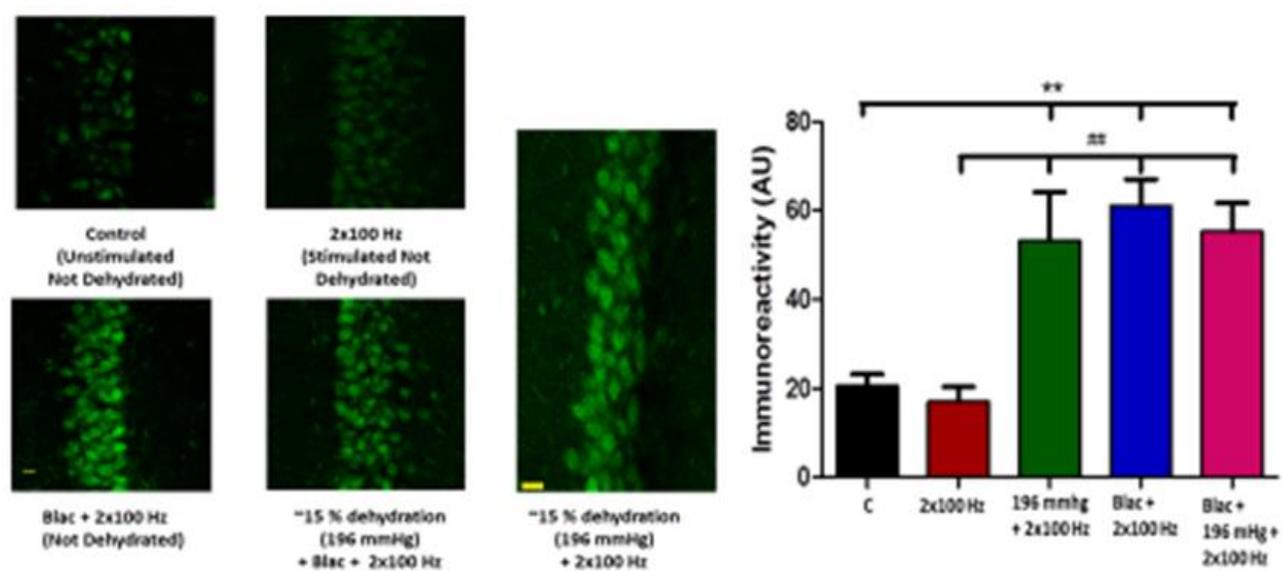

FIGURE 5

# Supplementary Material

*Main article title*

"INTERSTITIAL CELLS AND NEURONS RESPOND TO VARIATIONS IN HYDRATION"

# APPENDIX

## 1.1. The inert polymer, Polyethylene Glycol (PEG) is not cytotoxic and its effects are non-specific.

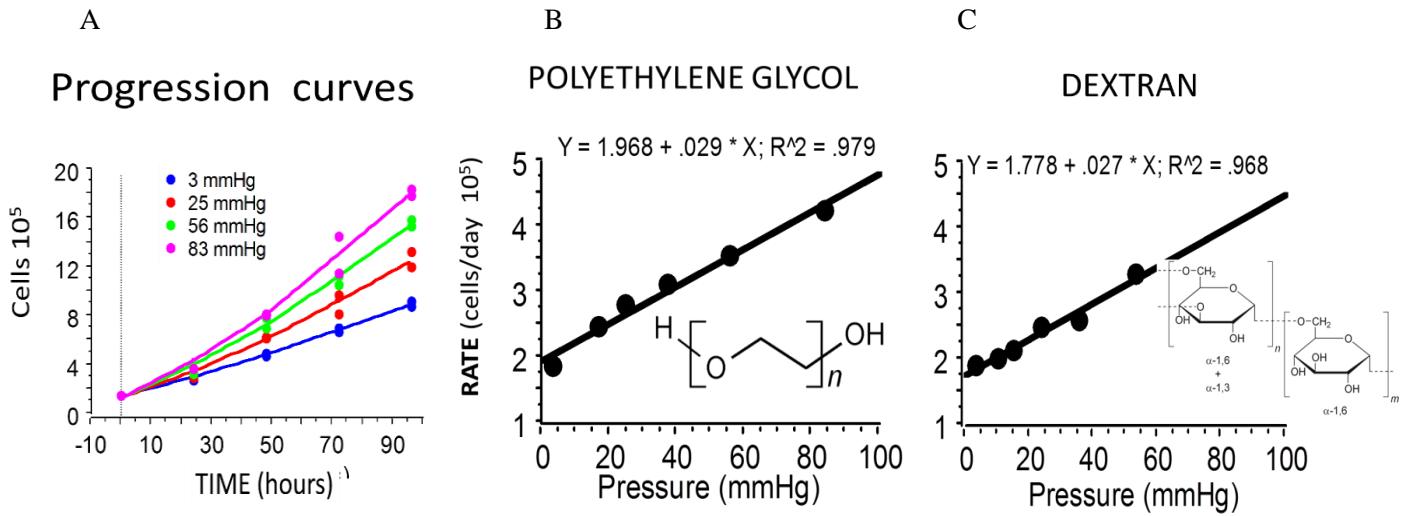

FIGURE 1a. *Cell growth-rate increases equally with pressure controlled using two distinct polymers.*

(A) Time trajectories of HL-60 cells numbers growing in suspension at the different colloidosmotic pressures indicated (pressure controlled with PEG); (B) growth rate as a function of pressure controlled with PEG and (C) growth rate as a function of pressure controlled with Dextran. Growth rates at each of 6 different pressure levels were approximated from linear regression analyses of cells numbers determined at 5 sequential times. Linear regression coefficient for both cell numbers versus time and rate versus pressure relationships were > 0.9 and had positive slopes with *P-values < 0.003*. Complete experimental protocols were reproduced n = 6 with PEG and n = 2 times with Dextran. The inserts in B and C illustrate the very distinct chemical formulas of the PEG and DEXTRAN polymer, respectively.

In the collagen gel contraction assays the water activity of the culture media was decreased to different levels with inert polymer; the differences in flux rate among levels inform about the impact of water activity gradients on cell responses. The results indicate that the gel contraction and the rate of fluid efflux from the gels were directly related to the water activity gradient which is inversely related to the activity of the polymer. However, because PEG is used at relatively high concentrations (0.02-0.1g/ml) to probe around physiological variations, we also tested for possible cytotoxicity of the polymer using the human promyelocytic leukemia cell line HL-60. These cells are anchorage-independent, sensitive to many toxins and grow fast in suspension which allows to determine cell viability and proliferation independently of adhesion and motility. Replicated cell suspensions (~1.5 X$10^5$ cells/ ml) were incubated during four days at 37 ◦C in 2 ml culture media (RPMI) with 2% fetal calf serum and inert polymer, either polyethylene glycol 8000 or Dextran T10 adjusted to colloidosmotic pressures ranging from 3 to 100 mmHg. Stock of Dextran solution had to be extensively dialyzed across a membrane with a 3000 MW cut off to remove small molecular weight chains and/or contaminants whereas the PEG solutions did not require dialysis.

Cell numbers increased with time and colloidosmotic pressure indicating that neither PEG nor Dextran solutions are cytotoxic. These results also largely exclude contributions from either specific polymer interactions or cell buoyancy changes. This is deduced from the fact that the two polymers have different chemical structure, their solutions have different densities and different amounts of polymer are required to reach equivalent pressures, leaving their nonspecific effect on water activity as the only ostensible common cause for the observed effect.

## *1.2. A note on osmotic pressure determination*.

In the present studies, the osmotic potential of the polymer solutions used were adjusted by reference to standard curves constructed by membrane osmometry with polymer solutions ranging between 0 and 10 % weight/weight in the tissue culture media. The osmometer was calibrated with a solution of human serum albumin as it is done to measure plasma colloidosmotic pressure clinically. To ensure that contributions from chains with molecular weights >3000 <8000, if present at significant concentrations, were included in the measured pressures we used a 3000 Mr cut-off membrane.

The values we obtained by membrane osmometry tend to be lower than those calculated from available published curves constructed using vapor pressure osmometry or thermocouple hygrometer techniques (1a). For example, in the only two other publications that we found on brain osmotic potential, the reported values for rat and cat brain were ~ 232 and ~ 578 mmHg (2.a, 3.a), respectively.

Using the curve in reference (1.a), the osmotic potential of a 5% solution of PEG 6000 comes to     ~ 400 mmHg whereas with the curves constructed by membrane osmometry and serum albumin calibration we found values ~ 100 mmHg. Therefore, the quantitative differences in the brain osmotic potential among the previous and our values can be largely ascribed to the methodology used to construct the reference curves. Although values in these previous studies seem unrealistically high from a physiologic point of view, the conclusion from them is qualitatively the same as in the present studies - that brain osmotic potential is much higher than that of plasma and cerebrospinal fluid.

## 1.3. *Note on intra and extra cytoplasmatic water activity*.

At steady state, the water in all tissues including the brain is distributed among cellular, vascular, and interstitial compartments. The role of specialized brain osmoreceptors in systemic vascular fluid volume regulation is well understood [14] and paravascular routes for fluid drainage in the brain are receiving renewed attention [18, 30, 29]. The water activity in cellular compartment appears to be very tightly regulated; solutes and hydrated surfaces densities are generally higher inside cells than outside, yet cell volumes remain constant.

There is extensive literature about mechanisms that maintain and restore volume, and these include both passive and energy dependent processes [4a]. Under isotonic conditions, passive mechanisms seem sufficient since in general cells suspended in chilled isotonic solutions do not noticeably swell indicating that cell volume can be maintain without metabolic energy expenditure and may harness forces arising from cytoskeletal elasticity [5a, 7a]. Most studies on cell volume changes, either under hyper or hypotonic conditions, use osmotic shifts that are very large compared, for example, to the physiologic and pathologic variations in plasma colloidosmotic pressure. Nevertheless, even after osmotic shocks exceeding plasma colloidosmotic pressure by order of magnitude, cell volume changes if any, are reversible [4.a].

Plasma colloidosmotic pressure, ~ 30 mmHg, is a very small fraction ~ 0.005 of the total osmotic pressure of physiologic fluids, ~5852 mmHg, therefore, its variations could be compensated by very small adjustments in crystalloids' concentration. [6.a].

The fact that the hydration potential that we measure in the tissues is generally about twice as large as the colloidosmotic pressure of plasma but less than the theoretical intracytoplasmic colloidosmotic pressure suggest that the intra and extra cellular compartments respond independently. Thus, observed cell responses to relatively minute changes in extracellular water activity- including increases in growth rate, as shown here for the HL-60 cells, decrease in gel-contraction rates by interstitial cells and the upregulation of transcription factor phosphorylation in neurons described in the main paper- represent activation of complex energy-dependent pathways not likely directly related to passive cell volume regulation.

How do cells sense minute fluctuation in the water activity of the extracellular spaces and program these complex responses is yet unknown, however, advances may come from investigating possible interactions between intracellular mechanics and molecular hydration transitions at membranes as proposed in the main article.

## 2. 1. Hydration parameters of dermis change after Heat-killing interstitial fibroblasts

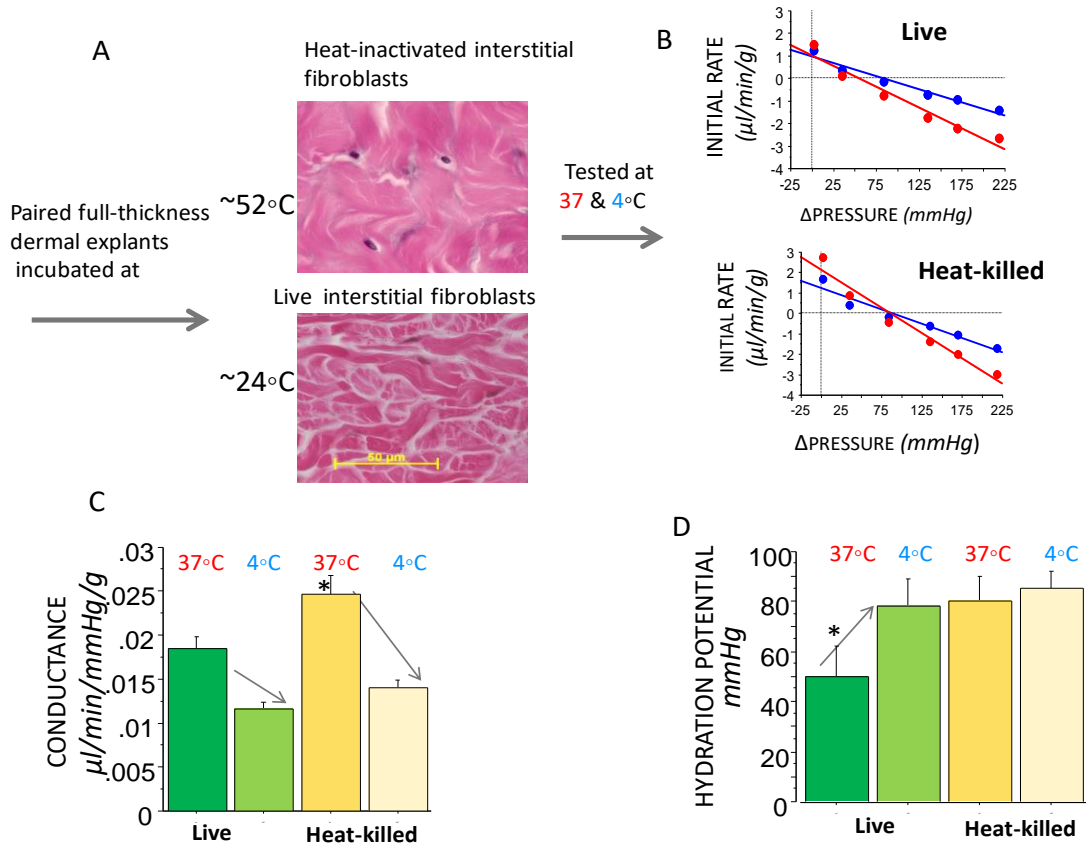

FIGURE 2a. *Heat inactivation of dermal explants, increased its fluid conductance and hydration potential relative to live explants at 37 °C (active cell metabolism) but not at 4 °C (metabolism inhibited) temperature.* Dermal explants were heated at 52°C to kill the cells without irreversibly denaturing the collagen matrix (panel A). Subsamples with either live or heat-killed cells were subjected to water activity titrations in baths incubated at either 4 or 37°C. Water activity was adjusted to increment the colloidosmotic pressure in the bath between 4 and 219 mmHg and the initial speed of fluid transfer between tissue and bath measured at each of 6 different water activity level (panel B) and plotted against the pressure increments. Hydration parameters, *Conductance* (panel C) and *Hydration potential* (panel D) were calculated from the slope and intercept of fitted lines. Statistically significant ($p<0.05$, $n = 11$) changes with temperature and between live and heat killed are indicated by arrows.

To explore individual contributions of cells and fiber- glycosaminoglycan gel to the hydration parameters of the interstitial matrix ideally, we would test them separately. In brain and myocardial tissue, it is not possible to preserve stromal structure and physiologic hydration while neutralizing interstitial cell-action, however, this ideal set-up can be approached in the dermal matrix due to its sturdy collagen- glycosaminoglycan weaved structure.

Matrix responses to hydration changes in the absence of fibroblast action were explored by comparing the hydration parameters of dermal samples heated to 52 °C to paired either non heated samples -or heated to 40 °C -where the fibroblasts remain alive. The temperature used to kill the interstitial fibroblasts induces little irreversible collagen denaturation [41] but eliminates all effects whether structural or mechanical mediated by the live fibroblasts network. Routine histologic examination demonstrated the expected effect of heat on cells displaying pyknotic nuclei and a cytoplasm that appears gelled and spilled over. The collagen fibers and bundles appeared swelled but not denatured. (Fig 3A).

In these experiments, both the samples heated to 52 °C and the control (paired samples keep at 24 °C) were tightly wrapped to prevent tissue water loss due to evaporation. Water content changed by < 2% and the change was not different in live and heat-inactivated samples [40]. After the heat treatment the tissues were equilibrated at 24 °C and then hydration parameters measured at either 4 or 37 °C.

Compared to live controls, there were significant differences in the hydration parameters of heat inactivated tissue Fig 3 B, C. The hydration potential decreased at physiologic temperatures in the live but not in the heat-killed dermis, again, suggesting hydration regulation by interstitial fibroblasts. These observations although consistent with an active role of the fibroblasts cannot rule out indirect effects due to the structural changes in the killed dermis. Further, the tissue compliance measured at 37 °C increased significantly in the heat killed dermis also suggesting that modulation of this parameter in live tissue includes both the material properties of the matrix as well as cell power.

Considered together, a plausible explanation compatible with the observed differences is that in live explants, cell action modulates the fluid-transfer properties of the matrix and helps maintaining its structural integrity. The fact that isolated cells incorporated in 3D collagen matrices also responded to changes in water activity within the biologically relevant variations measured in whole tissue explants is also consistent with a prominent role of contractile cells in regulating interstitial fluxes. In this regard, it is interesting to notice that normalized by weight, the power (i.e. fluid transfer work per unit of time) of live tissue explants and of cell in the 3D gels is similar, ~ 0.03 and ~0.05 µl/min/ mmHg/g, respectively. Although this could be a simple coincidence, it prompts us to speculate on whether there are programmed responses tuned to regulate local interstitial water activity within the narrow variations in hydration potentials measured in these studies.

# REFERENCES

## 1. References cited only in the appendix

## 2. References cited in both appendix and main article.

.